\documentclass[aps,prl,superscriptaddress,showpacs,twocolumn]{revtex4-1}
\usepackage{units}
\usepackage{amsmath}
\usepackage{amssymb}
\usepackage{graphicx}
\usepackage{bm}
\usepackage{multirow,color,relsize,ulem,microtype,mathtools}

\newcommand{\be}{\begin{equation}}
\newcommand{\ee}{\end{equation}}
\newcommand{\e}{\varepsilon}

\newcommand{\im}[1]{\text{Im}[#1]}

\newcommand{\gper}{\gamma_\perp}
\newcommand{\gpar}{\gamma_\|}

\begin{document}
\title{Condensation of Thresholds in Multimode Microlasers}

\author{Li Ge}
\affiliation{\textls[-18]{Department of Engineering Science and Physics, College of Staten Island, CUNY, Staten Island, NY 10314, USA}}
\affiliation{The Graduate Center, CUNY, New York, NY 10016, USA}
\author{Hui Cao}
\affiliation{Department of Applied Physics, Yale University, New Haven, CT 06520, USA}
\author{A. Douglas Stone}
\affiliation{Department of Applied Physics, Yale University, New Haven, CT 06520, USA}
\date{\today}

\begin{abstract}
We show from ab initio laser theory that by choosing an appropriate spatial pump profile, many different spatial modes of a typical microlaser can be turned on at the same pump energy, substantially increasing the number, $N$, of simultaneous lasing modes. The optimal pump profile can be obtained simply from knowledge of the space-dependent saturated gain profile when the system is uniformly pumped up to the $N$th modal threshold.
We test this general result by applying it to a two-dimensional diffusive random laser and a microdisk laser.
Achieving highly multimode lasing at reasonable pump powers is useful for reducing the spatial coherence of laser sources, making them suitable for use in speckle-free imaging and other applications.
\end{abstract}


\maketitle


The laser is a well studied driven-dissipative nonlinear system, and many aspects of the theory are well understood and tested experimentally \cite{Haken,Lamb}.  In the past two decades however, many new laser cavity designs have been introduced, both to study novel optical physics and in the search for efficient, on-chip microscale sources \cite{Microcavity1,Microcavity2,HuiReview}.  Unlike macroscopic laser cavities, where cavity design and intracavity components can be used to control the number of lasing modes, for microlasers modal control is less straightforward; moreover until recently there was no convenient theoretical approach to determining the number of lasing modes and their thresholds. Despite these challenges, modal control in microlasers offers a unique opportunity in regard to recent breakthroughs in speckle-free imaging \cite{speckleFree1,speckleFree2,speckleFree3}. While single-mode lasing is desirable in many applications, highly multimode lasing with spatially uncorrelated phases is a very convenient mechanism for reducing the spatial coherence of a bright laser source, allowing lasers to be used in full-field imaging microscopy and other applications requiring intense speckle-free sources. Compared with traditional low spatial coherence sources such as thermal lamps and light-emitting diodes, highly multimode lasers offer the advantage of higher power per mode, improved collection efficiency, and easier spectral control.

As has been known for some time \cite{Haken,HakenFu}, for essentially all microlasers, multimode lasing is stable due to spatial hole-burning: different spatial modes use distinct regions of the gain medium and can reach the lasing threshold (modal gain equals loss) at different pump strengths, despite saturation of the gain by modes which turn on earlier.  In addition, the large free spectral range in microlasers prevents population dynamics from effectively driving multimode laser instabilities. The modal thresholds of lasers in the absence of saturation are determined by two factors: the quality ($Q$) factor of the mode in the passive cavity, and the modal overlap with the gain medium, both spatially and spectrally.
Previous work on achieving highly multimode lasing in microlasers has focused on passive cavity engineering to create many modes with similar $Q$ factors, using, for example, random lasers \cite{speckleFree_RL} and chaotic lasers \cite{speckleFree_Chaotic}. This approach, however, usually leads to relatively low $Q$ factors and high thresholds.

In this letter we propose to exploit the spatial degrees of freedom of the pump to achieve highly multimode lasing in microlasers, which can be applied to both high-$Q$ and low-$Q$ cavities. The spatial pump profile can be controlled through spatial light modulators \cite{SLM}, phase masks, or eventually by multiple electrical contacts \cite{EC1,EC2,EC3}, and such an approach has been used empirically to achieve single-mode lasing \cite{Sebbah,Sebbah2,Seng_PRA15} and directional emission \cite{Hisch,Seng_APL14}. In contrast to these work using trial-and-error optimization to achieve modal control, here we show the existence of pump profiles leading to highly multimode lasing \textit{analytically}, using Steady-state Ab initio Laser Theory (SALT), a recently developed approach to predict the modal behavior in complex microlaser geometries with arbitrary pump profiles. SALT reduces the semiclassical laser equations to a set of {\it time-independent} self-consistent nonlinear wave equations that include the spatial hole burning effect exactly.
The SALT equations accurately find the solutions of the full semiclassical laser equations for nonuniformly pumped multimode  microlasers \cite{Science,TS,TSG,SPASALT,C-SALT}, and they have been used to predict new phenomena, such as re-entrant lasing near exceptional points \cite{EP_PRL}, which have since been observed \cite{Brandstetter,Peng};
Hence SALT (and approximations to it) have been used to study numerically modal control through variation of the pump profile.  However, due to the nonlinearity of the equations there have been no rigorous analytic results to guide these studies.

Here we present a very surprising result of this type. Suppose a laser cavity is pumped with some trial pump profile (e.g., uniform in space), and as the pump power increases, $N \gg 1$ modes turn on at thresholds $ D_0^{(\mu)}$ ($\mu = 1,2,\ldots N$).  We show that for any such trial profile there exists a refined pump profile which will cause all $N$ modes to turn on at the same ``master threshold" $D_m$, where $D_0^{(1)} < D_m < D_0^{(N)}$.  Above $D_m$ typically many more modes will lase for the same pump power as with the trial pump profile. We also show that this approach can be effective even given practical constraints on pump control.

Before going into the proof of threshold condensation and relevant examples, we note that degenerate thresholds arising from symmetry are well-known, e.g., counterpropagating modes in ring or disk lasers.  However, this case is only a pairwise degeneracy, and it is usually lifted by the nonlinearity, which randomly locks into one of the two possible states \cite{chiral1,chiral2,SA}.  There are also degenerate macroscopic cavity designs which have many modes with the same threshold.  Here, as noted, we are focusing on microlasers which can have {\it any} cavity design which supports many modes at high pump, and our approach requires no symmetry at all, nor any simple relationship among the condensed modes.

To understand why such a master threshold should exist, note that the pumped laser cavity itself is performing an optimization: as the pump increases, modes at different frequencies have different access to the gain, and all those which eventually lase have managed to balance gain and loss through positive feedback.  However in the nonlinear steady state, the lasing modes do not respond simply to the pump profile imposed externally; instead they respond to the saturated gain profile, which is strongly affected by the spatial variation and relative intensities of each mode.  Hence if one {\it imposes} an external pump at low power (so no modes lase) which follows spatially the {\it saturated gain profile} and simply increases the total pump power with this profile, at some power level all the modes will balance gain and loss with the now unsaturated gain susceptibility and start lasing together.

We now show that this simple argument is rigorously correct using the SALT equations \cite{Science,TS,TSG,SPASALT,C-SALT}, which find the steady-state solutions of the semiclassical Maxwell-Bloch equations \cite{Haken,Lamb} and $N$-level generalization thereof. SALT assumes a stationary population inversion in the gain medium (see the discussion in Ref.~\cite{directMethod}), which requires that the relaxation rate of the gain medium, $\gpar$, be small compared to the dephasing rate of the polarization, $\gper$, and the free spectral range of the laser, both satisfied for most microlasers. The high accuracy of SALT in this regime has been verified by comparing with time-dependent FDTD simulations \cite{OpEx08,N_level,directMethod,modeSwitching}. Although SALT, as well as the Maxwell-Bloch equations, treats homogeneously broadened gain media by construction, there is evidence that it applies qualitatively to certain inhomogeneously broadened gain media as well, such as InAs quantum dots lasers \cite{Seng_PRA15}.

The SALT equations  for
the steady-state lasing modes $\Psi_\mu(\vec{r};D_0)\,(\mu=1,\ldots,N)$ and their frequencies $\Omega_\mu$ take the
form \cite{directMethod}
\be
\left[ \nabla \times \nabla - [\e_c(\vec{r}) + \e_g(\vec{r};D_0)]\frac{\Omega_\mu^2}{c^2} \right]\Psi_\mu(\vec{r};D_0) = 0, \label{eq:SALT1}
\ee
where $c$ is the speed of light in vacuum and $\e_c(\vec{r})$ is the passive part of the cavity dielectric function independent of the pump strength, $D_0$. The electric field, $\Psi_{\mu}(\vec{r};D_0)$, is expressed in dimensionless form, measured in units $e_c = \hbar\sqrt{\gpar\gper}/2g$  where $g$ is the dipole matrix element of the lasing transition. The equations are to be solved with purely outgoing boundary conditions, and below the first threshold $D_0^{(1)}$ no solutions exist. Nontrivial solutions $\Psi_\mu(\vec{r};D_0)$ appear and increase in amplitude above each threshold $D_0^{(\mu)}$, and each $\Psi_\mu(\vec{r};D_0)$ oscillates at a real-valued lasing frequency $\Omega_\mu$, which in general varies with $D_0$.

Each mode interacts with itself and the other lasing modes via nonlinear gain saturation, which appears in the ``active" part of the dielectric function and takes the form \cite{SPASALT}:
\be
\hspace{-5mm}\mathclap{\e_g(\vec{r};D_0)\hspace{-2pt} =\hspace{-2pt} \frac{\gper}{\Omega_\mu \hspace{-2pt}-\hspace{-2pt} \omega_a \hspace{-1pt}+\hspace{-1pt} i\gper}\frac{D_0 f_0(\vec{r})}{1\hspace{-1pt}+\hspace{-1pt}\sum_{\nu=1}^N\Gamma_\nu|\Psi_\nu(\vec{r};D_0)|^2}.} \label{eq:epsg}
\ee
$\omega_a$ here is the atomic transition frequency, $\Gamma_\nu \equiv \gper^2/[\gper^2 + (\Omega_\nu-\omega_a)^2]$ is the Lorentzian gain curve evaluated at $\Omega_\nu$, and $f_0(\vec{r})\geq0$ is the externally imposed spatial profile of the pump, which we normalize by $\int_\text{cavity} f_0(\vec{r}) d\vec{r} = S$, where $S=\int_\text{cavity} d\vec{r}$ is the area of the cavity in two dimension (2D). As noted, the saturated gain profile depends strongly on the amplitude, spatial variation and frequency of the lasing modes, with the highest amplitude modes causing the most saturation.

Now consider $f_0(\vec{r})$, as the ``pilot" profile we wish to refine to cause all of the modes up to a target number, $N_t$, to lase at the same threshold, and assume we have solved the trial problem for $\Psi_\mu(\vec{r};D_0), \Omega_\mu$ up to the $N_t$th threshold, $D_t$.  As noted, for those $N_t$ modes the {\it saturated} gain profile balances gain and loss. Hence we take our refined pump profile to be proportional to the saturated gain profile
\be
{f}_m(\vec{r};D_t)=\frac{C(D_t)f_0(\vec{r})}{1+\sum_{\mu=1}^{N_t}\Gamma_\mu|\Psi_\mu(\vec{r};D_t)|^2},\label{eq:Transformation}
\ee
where $C(D_t)$ is a constant determined by the normalization $\int_\text{cavity} f_m(\vec{r};D_t) d\vec{r} = S$.
After replacing $f_0(\vec{r})$ by $f_m(\vec{r};D_t)$ in Eqs.~(\ref{eq:SALT1}) and (\ref{eq:epsg}), we insist that the {\it unsaturated} susceptibility with the new profile be identical to the saturated susceptibility of the original pilot problem at $D_t$. This requires that the appropriate pump value for the refined problem be
\be
{D}_m=\frac{D_t}{C(D_t)}, \label{eq:TH_transformed}
\ee
which uniquely determines a master threshold, at which all $N_t$ modes turn on under the new pump profile.
At $D_m$ all the lasing modes have the same spatial pattern and frequency as in the trial problem.  This construction was first noted in passing in Ref.~\cite{NMS}.

The above result is exact for the SALT equations, and therefore applies to many lasers, particularly microlasers as noted above.
However, while this result rigorously proves that all $N_t$ modes are at threshold at $D_m$, it does not prove that above $D_m$, all modes are lasing, and below $D_m$ none are. It is possible in principle that one or more of the modes have turned on at a lower pump value, acquired a negative slope before $D_m$, and stop lasing at $D_m$, which would be their ``off threshold."
However, while negative power slope due to modal interactions is possible, it requires special relationships between the modal profiles and lasing frequencies and typically also some optimization of the pump profile \cite{modeSwitching}. Since the modes involved here are arbitrary and the pump profile has been set by the requirement of degeneracy, it is highly unlikely that negative power slopes will occur at or near $D_m$ (see the Appendix).

\begin{figure}[t]
\centering
\includegraphics[width=\linewidth]{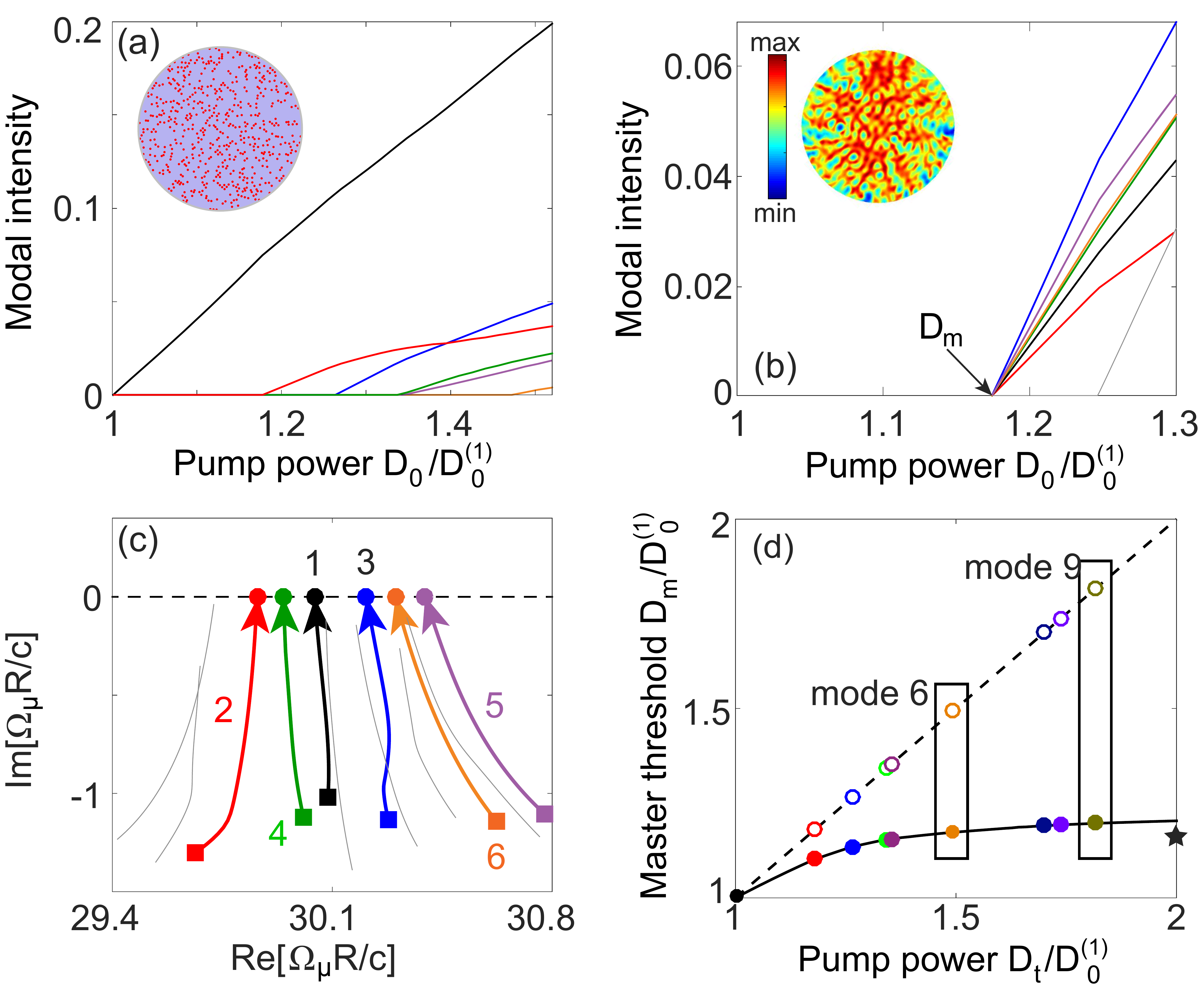}
\caption{(Color online) Condensation of multiple thresholds in a 2D random laser. (a) Modal intensity $I_\mu\equiv\int_\text{cavity} |\Psi_\mu(\vec{r};D_0)|^2 d\vec{r}/S$ as a function of pump power $D_0$ with uniform pumping. 
Inset: A disk pump of radius $R$ over an aggregate of scatterers with refractive index $n = 1.2$. The background index is 1. The gain medium is characterized by $\omega_aR/c=30$ and $\gper R/c=1$.
(b) Simultaneous onset of 6 modes above $D_m= 1.17D_0^{(1)}$ with the refined pump profile $f_m(\vec{r}; D_t=1.48D_0^{(1)})$. An additional mode (thin grey line) also turns on in the range of pump power shown. Inset: Refined pump profile and its color scale.
(c) The lasing frequencies of the 6 modes in (a) at $D_t$ are shown by filled circles, where the trajectories of the corresponding QB modes end up as the pump power increases from 0 (squares) to the master threshold $D_m$ with the refined pump profile.
The trajectories of 6 other QB modes that have not reached their thresholds are also shown (thin grey lines). The horizontal line indicates the threshold condition $\im{\Omega_\mu R/c}=0$. (d) $D_m$ as a function of $D_t$. Filled circles
show the values of $D_m$ when $N_t$ increases by 1, and open circles show the threshold of this new mode with uniform pumping. Two pairs of such values for $N_t=6$ and 9 are enclosed by boxes.
}\label{fig:2D}
\end{figure}

As a first demonstration of threshold condensation following this construction, we show in Fig.~\ref{fig:2D} the condensation of $N_t=6$ modes in a 2D diffusive random laser \cite{Science,RLreview1,RLreview2,RLreview3}, calculated via SALT.
(For 2D geometries the vector equation (\ref{eq:SALT1}) reduces to a nonlinear Helmholtz form \cite{SPASALT}).
Using a uniform pilot pump profile $f_0(\vec{r})=1$, we ramp up the pump power until 6 modes are lasing [see Fig.~\ref{fig:2D}(a)]. Expressed in terms of the 1st threshold $D_0^{(1)}$, the thresholds of the other 5 modes are 1.18, 1.27, 1.34, 1.35, and 1.48. Now using $D_t=1.48D_0^{(1)}$ and the corresponding saturated gain profile $f_m(\vec{r};D_t)$ as the refined pump profile, we find that the master threshold is given by $D_m=1.17D_0^{(1)}$. At this pump value only one mode lases with uniform pumping, while with the refined pump there are now six [see Fig.~\ref{fig:2D}(b)]. As a separate verification, we plot the trajectories of the corresponding resonance poles [quasi-bound (QB) mode frequencies] with increasing pump strength in Fig.~\ref{fig:2D}(c); the pump value at which a pole first reaches the real axis denotes the lasing threshold \cite{SPASALT}, and here all six poles reach the real axis simultaneously.

The same procedure can be applied to a larger target number of lasing modes in the same laser. For example, for this random laser the 9th mode starts lasing at $D^{(9)}_0=1.81D^{(1)}_0$ with uniform pumping. Now using this different value as $D_t$, and the different saturated gain profile as the refined pump profile, the master threshold is given by $D_m=1.20D^{(1)}_0$, a value where there are only two modes lasing with uniform pumping, whereas now there are nine.
Strikingly, the master threshold increases much more slowly with $N_t$ than does the $N_t$th threshold with uniform pumping [see Fig.~\ref{fig:2D}(d)]. For example, $D_m$ only increases by $0.03D_0^{(1)}$ when $N_t$ increases from 6 to 9, while the difference between the $6$th and $9$th thresholds with uniform pumping is $0.33D_0^{(1)}$.

The reason for this behavior is as follows: the ratio of $D_m$ and $D_t$ is given by the normalization constant $C(D_t)$ [Eq.~(\ref{eq:TH_transformed})]. By integrating both sides of Eq.~(\ref{eq:Transformation}) and using the normalization of $f_m(\vec{r};D_t)$, we find
\be
C(D_t)^{-1}=\int_\text{cavity} \frac{f_0(\vec{r})\,d\vec{r}}{1+\sum_{\mu=1}^{N_t}\Gamma_\mu|\Psi_\mu(\vec{r};D_t)|^2} .
\ee
On average, the saturation term in the denominator increases linearly with pump, so that (averaging over space)
$\sum_{\mu=1}^{N_t} \Gamma_\mu I_\mu(D_t) \approx a D_t/D_0^{(1)} - b$, (where $I_\mu(D_t)\equiv\int_\text{cavity} |\Psi_\mu(\vec{r};D_t)|^2 d\vec{r}/S$).    Thus $D_m \approx {D_t}/[1-b+a{D_t}/D_0^{(1)}] \rightarrow{D_0^{(1)}}/{a}$, when the pump power $D_t  \gg D_0^{(1)}$, indicating that $D_m$ remains of order $D_0^{(1)}$ even when $D_t$ becomes very large.  The asymptote of $D_m$ in Fig.~\ref{fig:2D}(d) is captured well by this approximation ($1.16D_0^{(1)}$; marked by the star).

\begin{figure}[t]
\centering
\includegraphics[width=\linewidth]{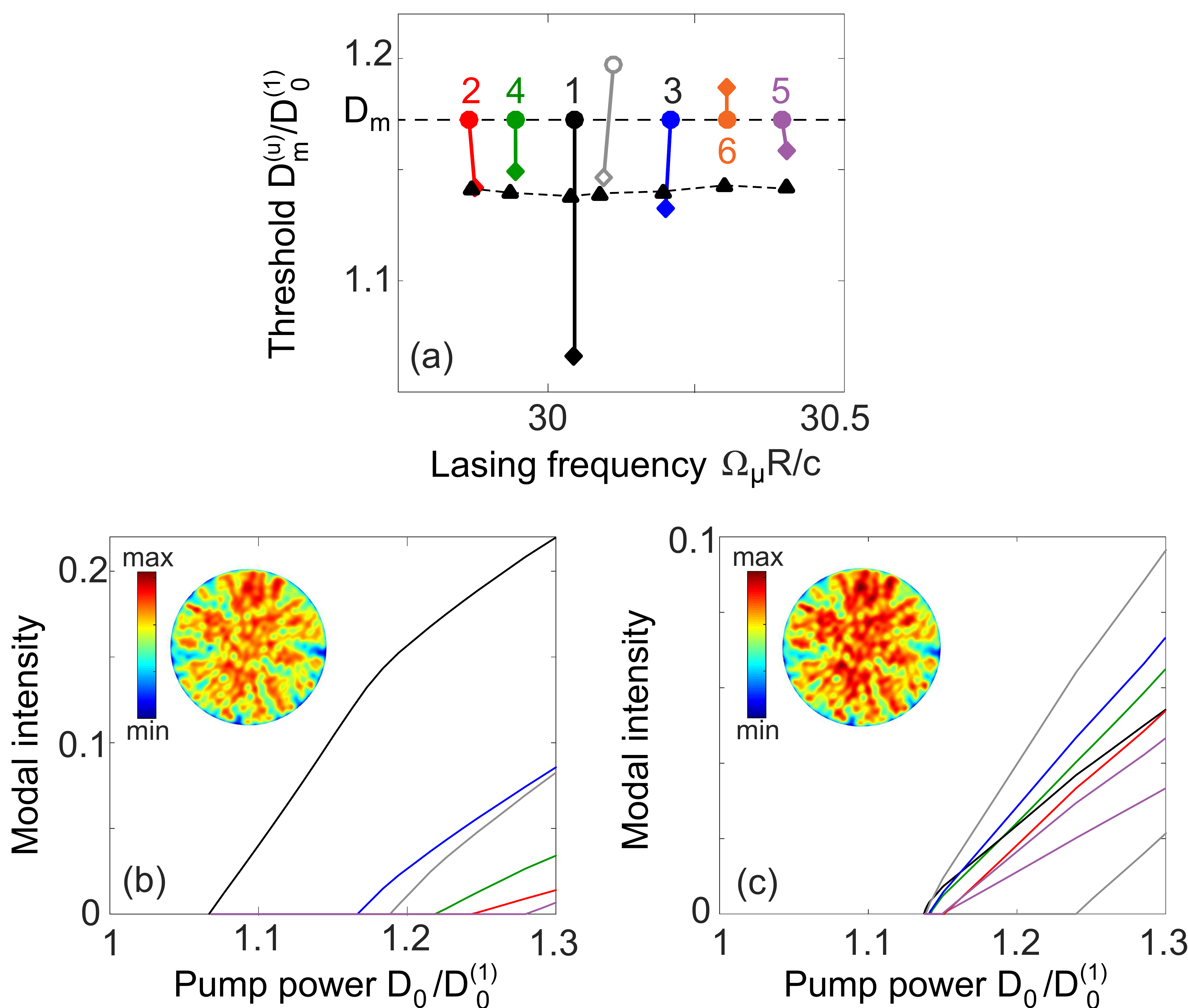}
\caption{(Color online) Effect of pump smearing and its correction. (a) Filled diamonds: lifting of the 6-fold threshold degeneracy (filled circles) in Fig.~\ref{fig:2D}(b) due to a Gaussian smoothing of the refined pump profile with $\sigma=R/40$ [see inset in (b)]. Open symbols show an additional mode. Connected black triangles: restoration to a quasi-degeneracy after an ``error correction" procedure with modal-intensity changes $\Delta I_{1-7}(D_t)=0.82,0.01,0.25,0.08,-0.05,-0.10,0.13$.
Modal interactions under the smeared pump profiles are neglected in calculating the degeneracy-lifted thresholds $D_m^{(u)}$ but included in (b) and (c). (b) Modal intensity $I_\mu$ as a function of pump power $D_0$ with $\bar{f}_m(\vec{r};D_t)$ before ``error correction." (c) Same as (b) but after ``error correction."
}\label{fig:blur}
\end{figure}

The saturated gain profile used to generate the refined pump typically varies on the scale of the wavelength of light in the cavity, whereas approaches to shape the pump profile mentioned in the introduction will have limited resolution due to the diffraction limit, carrier diffusion and other effects.  In addition, the saturated gain profile is typically not directly measurable but must be calculated from some model of the cavity, and will be subject to corresponding inaccuracies. Thus there will be limits on our ability to generate the ideal pump profile leading to exact degeneracy. To estimate this effect, we perform a Gaussian smoothing of the refined pump profile used in Fig.~\ref{fig:2D}:
\be
\bar{f}_m(\vec{r};D_t)  = \frac{\bar{C}(D_t)}{2\pi\sigma^2}\int_\text{cavity} \,f_m(\vec{\varsigma};D_t)\, e^{-\frac{(\vec{r}-\vec{\varsigma})^2}{2\sigma^2}} d\vec{\varsigma},
\ee
where $\bar{C}(D_t)$ is a normalization constant similar to $C(D_t)$. We note that a quite noticeable reduction of pump details already takes place at $\sigma=R/40$ [see the inset in Fig.~\ref{fig:2D}(b)].
Nevertheless, we still find a significant enhancement of multimode lasing [see Fig.~\ref{fig:blur}(b)]: while the 6th threshold is now at $1.27D_0^{(1)}$ and higher than the master threshold ($1.17D_0^{(1)}$), it is still much lower than its value with uniform pumping ($1.48D_0^{(1)}$). $D_0^{(1)}$ here refers the first threshold under the pilot pump profile.

In addition, we can perform further optimization by treating the refined pump profile itself as a variational function, and using the intensities in the denominator of Eq.~(\ref{eq:Transformation}) as variational parameters to compensate much of the threshold splitting.
For example, the dominant effect of pump smearing in Fig.~\ref{fig:blur}(a) is a much lower threshold of mode 1. To reverse this change, we increase the suppression of mode 1 in the refined pump profile (before smearing) by increasing the intensity $|\Psi_1(\vec{r};D_t)|^2$ (and $I_1(D_t)$). Due to cross-saturation, this treatment also changes the thresholds of the other modes but typically to a lesser extent. Hence by adjusting each modal intensity in the appropriate direction to compensate its splitting due to smearing, quasi-degeneracy can be restored as shown in Figs.~\ref{fig:blur}(a) and (c).  Note that the quasi-degenerate master threshold after ``error correction" is given by $\bar{D}_m\approx1.15D_0^{(1)}$, which is even lower than the degenerate one.
This illustrates the point that $D_m$ is not a lower bound on the pump value where $N_t$ modes can lase; but it is an excellent
starting point for optimization.

\begin{figure}[t]
\centering
\includegraphics[width=\linewidth]{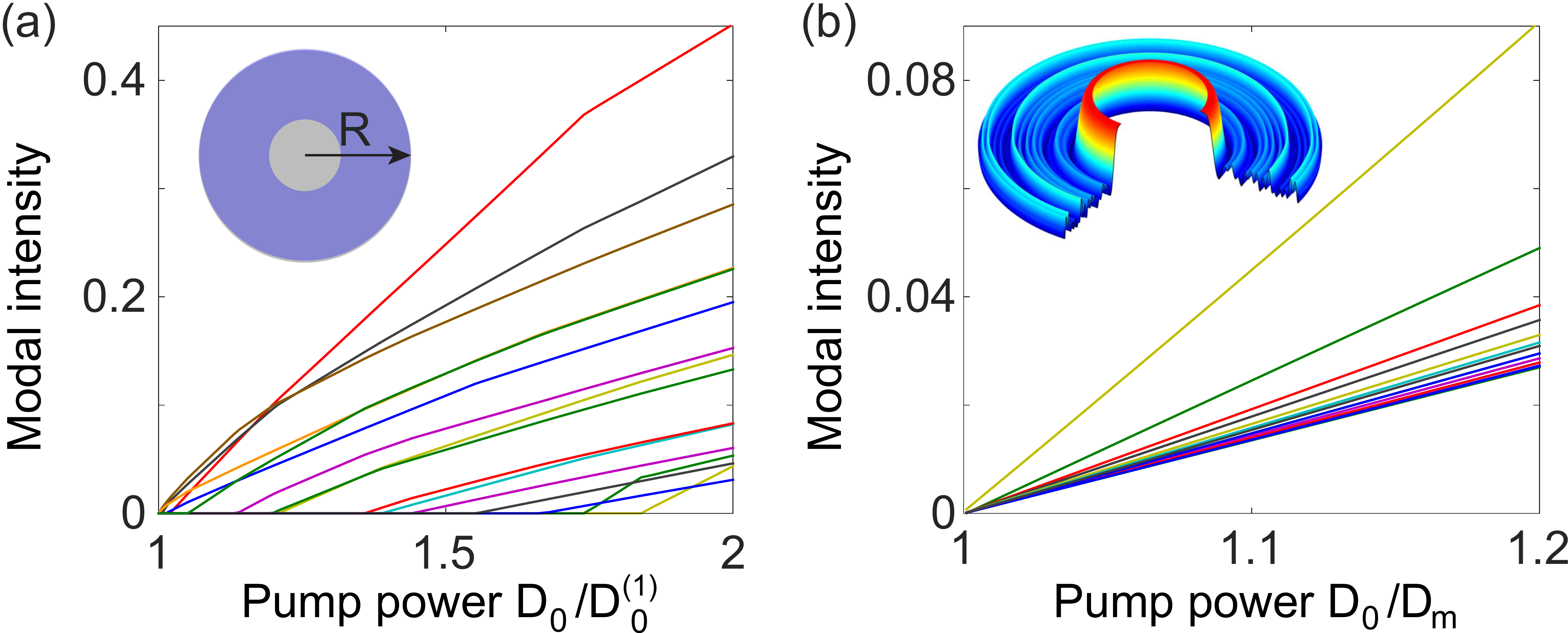}
\caption{(Color online) Condensation of multiple thresholds in a microdisk laser of radius $R$. (a) Modal intensity $I_\mu$ as a function of pump power $D_0$ with a ring-shaped pilot pump profile (inset). The 16th mode starts lasing at $D_0=1.83D^{(1)}_0$, and $\gper R/c=0.5$ is used. (b) Simultaneous onset of all 16 modes above $D_m=1.08D^{(1)}_0$ with the refined pump profile $f_m(\vec{r}; D_t=1.83D^{(1)}_0).$
The 3D rendering of the latter is shown as the inset, one quarter of which is removed to show its radial profile.}\label{fig:disk}
\end{figure}

As a final example, we apply the threshold condensation procedure to a microdisk laser (see Fig.~\ref{fig:disk}), a well studied multimode microlaser of technological interest.  We choose a high index contrast typical of semiconducting devices although we assume an atomic-like gain medium (e.g. quantum dots). The relevant electromagnetic modes are high-$Q$ whispering-gallery modes (WGMs), confined by near total internal reflection. For the reason discussed already in the introduction, we only consider WGMs of one symmetry, e.g., the clockwise rotation (positive azimuthal quantum number $m$), which preserves the rotational symmetry of the system when gain saturation is considered. With $n=3.3+10^{-4}i$ and $\omega_a R/c=30$, this microdisk laser supports WGMs of $m$ up to 100. It is natural here to choose a non-uniform pilot profile, and we take a ring-shaped pump profile ($f_0(\vec{r})=0$ for $r<R/3$).  With this choice
the 1st mode has $m=80$ and the 16th mode of $m=37$ starts lasing at $D_0^{(16)}=1.84D_0^{(1)}$ [see Fig.~\ref{fig:disk}(a)]. Using this pump value as $D_t$, the master threshold occurs at $D_m=1.08D_0^{(1)}$, beyond which all 16 modes start lasing simultaneously [see Fig.~\ref{fig:disk}(b)].

In summary, we have shown that the spatial hole-burning nonlinearity of a laser can be utilized to refine the pump profile, resulting in the  simultaneous lasing of many modes at relatively low pump power. This analytic property of the lasing equations can be used as a guide to find spatial profiles leading to control of multimode lasing properties, even if there are limitations on the spatial resolution of the pump. L.G. acknowledges partial support under PSC-CUNY Grant No. 68698-0046 and NSF Grant No. DMR-1506987. H.C. acknowledges support under NSF Grant No. DMR-1205307. A.D.S. acknowledge support under NSF Grant No. DMR-1307632.

\appendix
\section{Appendix: Mode Suppression Beyond the Master Threshold}
With a refined pump profile $f_m(\vec{r};D_t)$, we have shown in the main text that all $N_t$ modes lase simultaneously once the pump power is increased beyond the resulting master threshold $D_m$. There are two rare scenarios where this behavior breaks down, with one or more of the $N_t$ mode suppressed beyond $D_m$. We discuss these two scenarios in this appendix.

\begin{figure}[b]
\centering
\includegraphics[width=\linewidth]{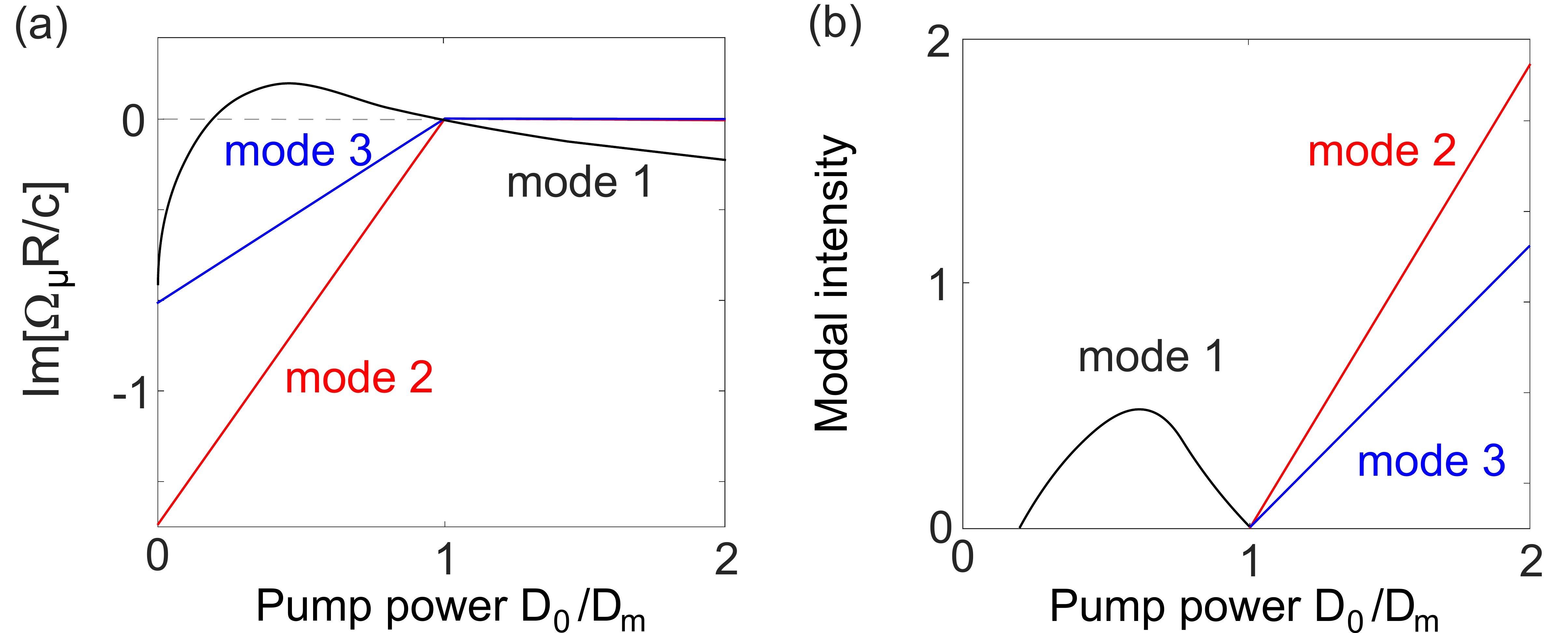}
\caption{(Color online) Schematics showing that the master threshold in principle can be where a mode is turned off instead of turned on. $D_m$ in this case is the ``on" threshold of mode 2 and 3 and ``off" threshold of mode 1.}\label{fig:double_crossing}
\end{figure}

The first scenario in which one or more of the $N_t$ modes could in principle be suppressed beyond their condensed threshold $D_m$ is a linear effect, where the trajectory of the corresponding quasi-bound mode frequency $\Omega_\mu(D_0)$ crosses the real axis at $D_0=D_m$ from above. This behavior is caused by an exceptional point \cite{unconventional,EP_PRL,EP_CMT}, and such an $\Omega_\mu(D_0)$ has another crossing with the real axis at a lower pump power (see the schematics in Fig.~\ref{fig:double_crossing}). We have not find a case where this rare scenario takes place, which requires fine tuning of the system parameters close to an exceptional point.

The second scenario is a nonlinear effect. It occurs if the onset of mode competition above $D_m$ would lead to a negative power slope of one or more modes. To study this scenario analytically, we assume the cavity has a high $Q$ factor and consider first the $N_t=2$ case for simplicity. For a high-$Q$ cavity SALT can be approximated by a simple set of intensity equations \cite{SPASALT}, which take the following form here:
\be
\frac{D_0}{D_m} - 1 = \chi_{11}\mathfrak{I}_1 + \chi_{12}\mathfrak{I}_2 = \chi_{22}\mathfrak{I}_2 + \chi_{21}\mathfrak{I}_1.
\label{eq:SPASALT}
\ee
Although $\mathfrak{I}_\mu \equiv \Gamma_\mu |\int_\text{cavity} f_m(\vec{r}) \Psi_\mu(\vec{r};D_0)^2 d\vec{r}/S|$ is defined differently from the modal intensity $I_\mu$ introduced in the main text, they are proportional to each other in a high-$Q$ cavity, and the normalized spatial mode profile $\varphi_\mu(\vec{r})=\Psi_\mu(\vec{r};D_0)\sqrt{\Gamma_\mu/\mathfrak{I}_\mu}$ is approximately real and varies little above threshold.
$\chi_{\mu\nu}=\left|\int_\text{cavity} f_m(\vec{r})\varphi_\mu(\vec{r})^2|\varphi_\nu(\vec{r})|^2 d\vec{r}/S\right|$ gives the self-interaction coefficients when $\mu=\nu$ and the cross-interaction coefficients when $\mu\neq\nu$, and we note that $\chi_{21}\approx\chi_{12}$ holds in a high-$Q$ cavity. It is straightforward to see that a negative power slope for either $\mathfrak{I}_1$ or $\mathfrak{I}_2$ requires
\be
\frac{\chi_{22}-\chi_{12}}{\chi_{11}-\chi_{21}}=\frac{\mathfrak{I}_1}{\mathfrak{I}_2}<0. \label{eq:ineq}
\ee
In other words, the cross-interaction coefficients $\chi_{12},\chi_{21}$ need to be larger than one of the self-interaction coefficients and smaller than the other. This condition is again very rare in a high-$Q$ cavity which features $\chi_{11},\chi_{22}\gg\chi_{12},\chi_{21}$ in general; only recently was the condition (\ref{eq:ineq}) reported that leads to interaction-induced mode switching (IMS) \cite{modeSwitching}. Using the multimode form of Eq.~(\ref{eq:SPASALT}), the criterion (\ref{eq:ineq}) can be generalized to $N_t>2$ cases straightforwardly, i.e.
\be
\mathfrak{I}_\mu \propto \sum_{\nu=1}^{N_t} [\chi^{-1}]_{\mu\nu}<0 \label{eq:ineq2}
\ee
for one or more of the $N_t$ modes, where $\chi^{-1}$ is the inverse matrix of $\chi$.

\begin{figure}[h]
\centering
\includegraphics[width=\linewidth]{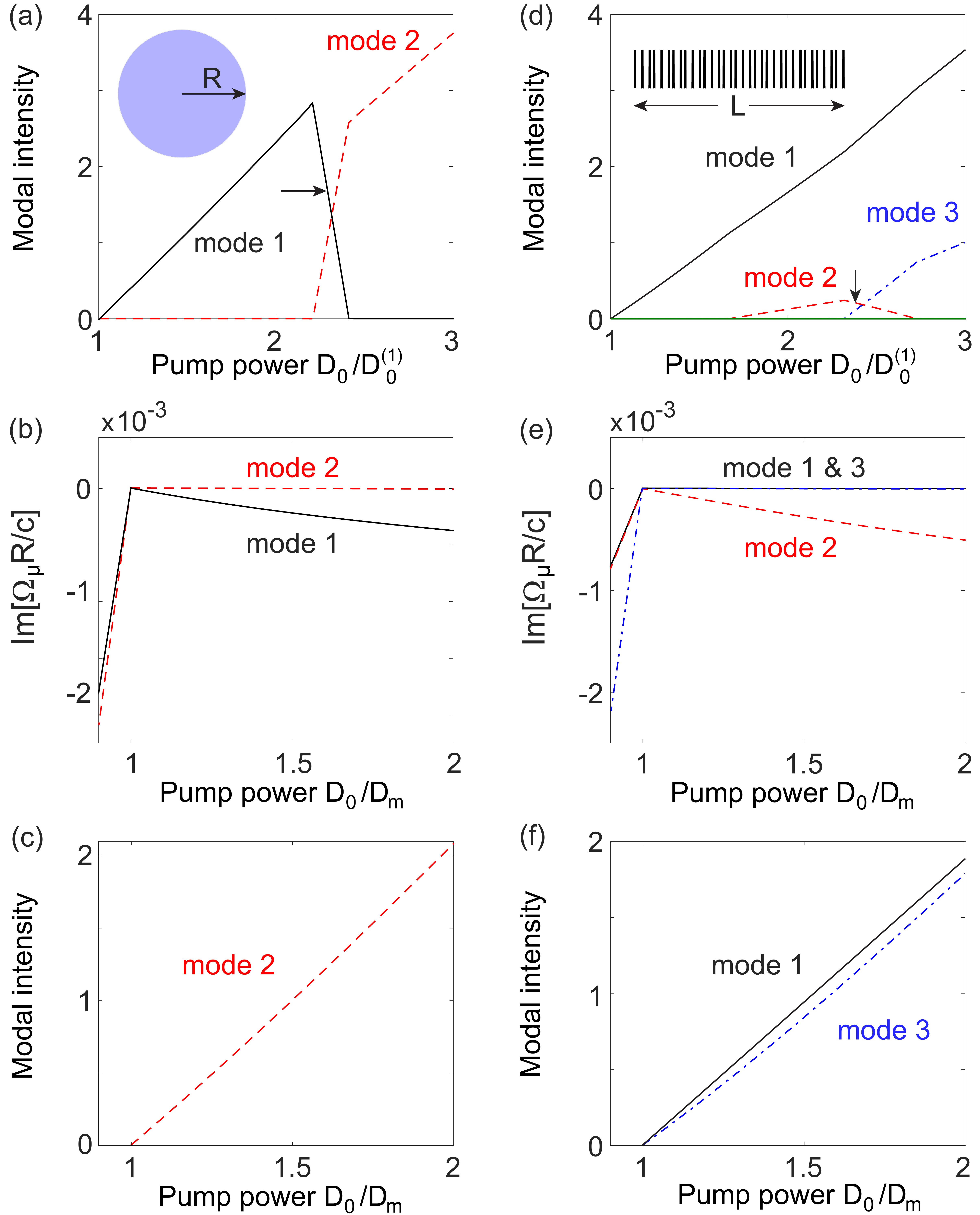}
\caption{(Color online) Mode suppression beyond the master threshold in a microdisk laser (a-c) and an aperiodic laser (d-f).
(a) and (d) Modal intensity $I_\mu$ as a function of pump power $D_0$ with uniform pumping. Both cases feature an interaction-induced mode switching (IMS), for the 1st and 2nd modes respectively. The arrows point to their negative power slopes at $D_t=2.29D^{(1)}_0$ in (a) and $D_t=2.38D^{(1)}_0$ in (c).
Insets: Schematics of a microdisk laser with $n_c=2+0.01i$, $\omega_aR/c=4.83$ and $\gper R/c=1$ in (a) and an aperiodic laser with dielectric layers of $n_c=1.5$ (in air), $\omega_aR/c=122.51$ and $\gper R/c=17.46$ in (d).
(b) and (e) $\im{\Omega_\mu R/c}$ as a function $D_0$ with the refined pump profile $f_m(\vec{r}; D_t)$.
$N_t=2,3$ and $D_m/D^{(1)}_0=1.54,1.37$ in (b) and (e), respectively.
(c) and (f) Same as (a) and (d) but with their respective $f_m(\vec{r}; D_t)$.}\label{fig:IMS}
\end{figure}

In Fig.~\ref{fig:IMS} we apply the condensation procedure to the microdisk laser studied in Ref.~\cite{modeSwitching} that exhibits IMS: the 2nd mode turns on at $D^{(2)}_0=2.21D^{(1)}_0$ before it switches off the 1st mode via a negative power slope at $D_0=2.41D^{(1)}_0$ [see Fig.~\ref{fig:IMS}(a)]. We choose $D_t=2.29D^{(1)}_0$ in the cross-over region, which leads to $D_m=1.54D^{(1)}_0$. As Fig.~\ref{fig:IMS}(b) shows, $\Omega_{1,2}$ reach the real axis simultaneously as the pump power is ramped up to $D_m$ with the pump profile $f_m(\vec{r};D_t)$. But as soon as $D_0$ goes beyond $D_m$, $\Omega_1$ is forced into the lower half of the complex plane again due to mode competition, leading to single mode lasing of the 2nd mode only [see Fig.~\ref{fig:IMS}(c)].

While Eq.~(\ref{eq:SPASALT}) and the criterion (\ref{eq:ineq2}) do not apply to low-$Q$ cavities, the correlation between IMS and the suppression of certain mode(s) beyond $D_m$ still seems to hold. In Figs.~\ref{fig:IMS}(d-f) we show such an example in a one-dimensional (1D) aperiodic cavity. The 2nd mode exhibits a negative power slope after the onset of the 3rd mode at $D_0=1.66D^{(1)}_0$, and it is switched off at $D_0=2.73D^{(1)}_0$. When taking any pump value in this range as $D_t$, we find that the 2nd mode is suppressed beyond $D_m$ with the refined pump profile $f_m(\vec{r};D_t)$.

\begin{figure}[t]
\centering
\includegraphics[width=\linewidth]{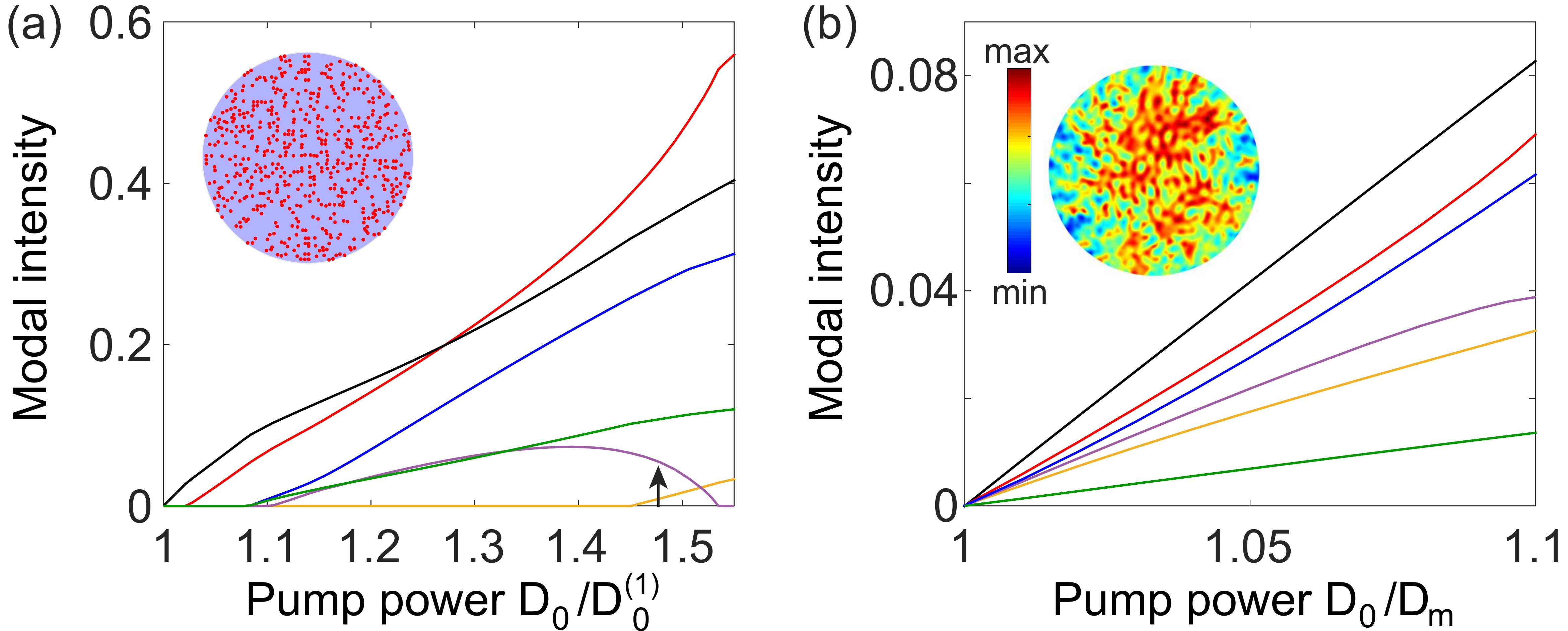}
\caption{(Color online) Condensation of multiple thresholds in a 2D random laser. (a) Modal intensity $I_\mu$ as a function of pump power $D_0$ with uniform pumping. The arrow points to the negative power slope of the 5th mode at $D_t=1.48D^{(1)}_0$ chosen for $N_t=6$. Inset: Schematics of the 2D random laser. The parameters are the same as in Fig.~1 of the main text except for a different disorder. (b) Simultaneous onset of all 6 modes above $D_m=1.35D^{(1)}_0$ with the refined pump profile $f_m(\vec{r}; D_t)$. The latter is shown in the inset with its color scale. }\label{fig:2D2}
\end{figure}

We note that IMS does not only feature a negative power slope; this negative power slope must be induced suddenly by the onset of a new lasing mode. If the negative power slope is due to mode mixing \cite{Science}, we find that all $N_t$ modes still lase simultaneously above $D_m$. One such example is shown in Fig.~\ref{fig:2D2} for a 2D random laser.

\bibliographystyle{longbibliography}

\begin{thebibliography}{99}

\bibitem{Lamb} M. Sargent III, M. O. Scully, and W. E. Lamb, Jr., \textit{Laser Physics} (Addison-Wesley, 1974).
\bibitem{Haken} H.~Haken, \textit{Light: Laser Dynamics} (North-Holland Physics Publishing, 1985), Vol. II.

\bibitem{Microcavity1} R. K. Chang and A. J. Campillo, \textit{Optical Processes in Microcavities} (World Scientific, 1996).
\bibitem{Microcavity2} K. J. Vahala, \textit{Optical Microcavities} (World Scientific, 2004).
\bibitem{HuiReview} H. Cao and J. Wiersig, ``Dielectric microcavities: Model systems for wave chaos and non-Hermitian physics,” Rev. Mod. Phys. \textbf{87}, 61 (2015).



\bibitem{speckleFree1} M. Nixon, B. Redding, A. A. Friesem, H. Cao, and N. Davidson, ``Efficient method for controlling the spatial coherence of a laser," Opt. Lett. \textbf{38}, 3858 (2013).
\bibitem{speckleFree2} B. Redding, P. Ahmadi, V. Mokan, M. Seifert, M. A. Choma, and H. Cao, ``Low-spatial-coherence high-radiance broadband fiber source for speckle free imaging," Opt. Lett. \textbf{40}, 4607 (2015).
\bibitem{speckleFree3} S. Knitter, C. Liu, B. Redding, M. K. Khokha, M. A. Choma, H. Cao. ``Coherence switching of a degenerate VECSEL for multimodality imaging," Optica \textbf{3}, 403 (2016).

\bibitem{HakenFu} H. Fu and H. Haken, ``Multifrequency operations in a short-cavity standing-wave laser," Phys. Rev. A \textbf{43}, 2446 (1991).

\bibitem{speckleFree_RL} B. Redding B, M. A. Choma, and H. Cao, ``Speckle-free laser imaging using random laser illumination," Nature Photon. \textbf{6}, 355 (2012).
\bibitem{speckleFree_Chaotic} B. Redding, A. Cerjan, X. Huang, M. L. Lee, A. D. Stone, M. A. Choma, and H. Cao, ``Low spatial coherence electrically pumped semiconductor laser for speckle-free full-field imaging,” Proc. Nat. Acad. Sci. \textbf{112}, 1304 (2015).




\bibitem{SLM} M. Leonetti and C. Lopez, ``Active subnanometer spectral control of a random laser," Appl. Phys. Lett. {\bf 102}, 071105 (2013).

\bibitem{EC1} T. Fukushima, T. Harayama, P. Davis, P. O. Vaccaro, T. Nishimura, and T. Aida, ``Ring and axis mode lasing in quasi-stadium laser diodes with concentric end mirrors," Opt. Lett. \textbf{27}, 1430 (2002).
\bibitem{EC2} M. Kneissl, M. Teepe, N. Miyashita, N. M. Johnson, G. D. Chern, and R. K. Chang, ``Current-injection spiral-shaped microcavity disk laser diodes with unidirectional emission," Appl. Phys. Lett. \textbf{84}, 2485 (2004).
\bibitem{EC3} S. Shinohara, T. Harayama, T. Fukushima, M. Hentschel, T. Sasaki, and E. E. Narimanov, ``Chaos-assisted directional light emission from microcavity lasers," Phys. Rev. Lett. \textbf{104}, 163902
(2010).



\bibitem{Sebbah} N. Bachelard, J. Andreasen, S. Gigan, and P. Sebbah, ``Taming random lasers through active spatial control of the pump," Phys. Rev. Lett. {\bf 109}, 033903 (2012).
\bibitem{Sebbah2} N. Bachelard, S. Gigan, X. Noblin, and P. Sebbah, ``Adaptive pumping for spectral control of random lasers,” Nature Phys. \textbf{10}, 426 (2014).
\bibitem{Seng_PRA15} S. F. Liew, L. Ge, B. Redding, G. S. Solomon, and H. Cao, ``Pump-controlled modal interactions in microdisk lasers,"
    Phys. Rev. A \textbf{91}, 043828 (2015).

\bibitem{Hisch} T. Hisch, M. Liertzer, D. Pogany, F. Mintert, and S. Rotter, ``Pump-controlled directional light emission from
random lasers,” Phys. Rev. Lett. \textbf{111}, 023902 (2013).
\bibitem{Seng_APL14} S. F. Liew, B. Redding, L. Ge, G. S. Solomon, and H. Cao, ``Active control of emission directionality of semiconductor microdisk lasers," App. Phys. Lett. \textbf{104}, 231108 (2014).


\bibitem{Science} H. E. T\"ureci, L. Ge, S. Rotter, and A. D. Stone, ``Storng interactions in multimode random lasers," Science {\bf 320}, 643 (2008).
\bibitem{TS} H.~E.~T\"ureci, A.~D.~Stone and B.~Collier, ``Self-consistent multimode lasing theory for complex or random lasing media,"
    Phys.~Rev.~A \textbf{74}, 043822 (2006).
\bibitem{TSG} H.~E.~T\"ureci, A.~D.~Stone and L. Ge, ``Theory of the spatial structure of nonlinear lasing modes,"
Phys.~Rev.~A \textbf{76}, 013813 (2007).
\bibitem{SPASALT} L.~Ge, Y.~D.~Chong, and A. D. Stone, ``Steady-state ab initio laser theory: generalizations and analytic results," Phys. Rev. A {\bf 82}, 063824 (2010).
\bibitem{C-SALT} A. Cerjan, Y. D. Chong and A. D. Stone, ``Steady-state ab initio laser theory for complex gain media," Opt. Express \textbf{23}, 6455 (2015).

\bibitem{EP_PRL} M.~Liertzer, L.~Ge, A.~Cerjan, A.~D.~Stone, H.~E.~T\"{u}reci, and S.~Rotter, ``Pump-induced exceptional points in lasers," Phys. Rev. Lett. {\bf 108}, 173901 (2012).
\bibitem{Brandstetter} M. Brandstetter, M. Liertzer, C. Deutsch, P. Klang, J. Sch\"oberl, H. E. T\"ureci, G. Strasser, K. Unterrainer, and S. Rotter, ``Reversing the pump dependence of a laser at an exceptional point," Nature Comm. \textbf{5}, 4034 (2014).
\bibitem{Peng} B. Peng, Ş. K. \"Ozdemir, S. Rotter, H. Yilmaz, M. Liertzer, F. Monifi, C. M. Bender, F. Nori, and L. Yang, ``Loss-induced suppression and revival of lasing,” Science \textbf{346}, 328 (2014).

\bibitem{chiral1} Q.-T. Cao, H.-M. Wang, C.-H. Dong, H. Jing, R.-S. Liu, X. Chen, L. Ge, Q. Gong, and Y.-F. Xiao, ``Experimental demonstration of spontaneous chirality in a nonlinear microresonator," arXiv:1607.01459 (2016).
\bibitem{chiral2} L. Del Bino, J. M. Silver, S. L. Stebbings, and P. Del’Haye, ``Symmetry breaking of counter-propagating light in a nonlinear resonator," arXiv:1607.01194 (2016).
\bibitem{SA} S.~Burkhardt, M.~Liertzer, D.~O.~Krimer, S.~Rotter, ``Steady-state ab-initio laser theory for lasers with fully or nearly degenerate resonator modes," Phys. Rev. A \textbf{92}, 013847 (2015).

\bibitem{directMethod} S. Esterhazy, D. Liu, M. Liertzer, A. Cerjan, L. Ge, K. G. Makris, A. D. Stone, J. M. Melenk, S. G. Johnson, and S. Rotter, ``Scalable numerical approach for the steady-state ab-initiolaser theory," Phys.~Rev.~A \textbf{90}, 023816 (2014).
\bibitem{OpEx08} L. Ge, R. J. Tandy, A. D. Stone, and H. E. T\"ureci, ``Quantitative verification of ab initio self-consistent laser theory," Opt. Express {\bf 16}, 16895 (2008).
\bibitem{N_level} A. Cerjan, Y. D. Chong, L. Ge, and A. D. Stone, ``Steady-State ab initio laser theory for N-level lasers,"
Opt. Express \textbf{20}, 474 (2012).
\bibitem{modeSwitching} L. Ge, D. Liu, A. Cerjan, S. Rotter, H. Cao, S. G. Johnson, H. E. T\"ureci, and A. D. Stone, ``Interaction-induced mode switching in steady-state microlasers," Opt. Express \textbf{24}, 41 (2016).

\bibitem{NMS} L. Ge, ``Selective excitation of lasing modes by controlling modal interactions," Opt. Express \textbf{23}, 30049 (2015).

\bibitem{RLreview1} H. Cao, ``Review on latest developments in random lasers with coherent feedback," J. Phys. A: Math. Gen. \textbf{39}, 467 (2005).
\bibitem{RLreview2} D. S. Wiersma, ``The physics and applications of random lasers,'' Nature Phys. \textbf{4}, 359 (2008).
\bibitem{RLreview3} J. Andreasen {\it et al.} ``Modes of random lasers," Adv. Opt. Photon. {\bf 3}, 88 (2011).

\bibitem{unconventional} L. Ge, Y. D. Chong, S. Rotter, H. E. T\"ureci, and A. D. Stone, ``Unconventional modes in lasers with spatially varying gain and loss," Phys. Rev. A \textbf{84}, 023820 (2011).
\bibitem{EP_CMT} R. El-Ganainy, M. Khajavikhan, and L. Ge, ``Exceptional points and lasing self-termination in photonic molecules,"
Phys. Rev. A \textbf{90}, 013802 (2014).


\end{thebibliography}

\end{document}